\documentclass[preprint,12pt]{elsarticle}
 \date{}

\usepackage{amssymb}
\usepackage[english]{babel}
\usepackage{comment}

\usepackage[letterpaper,top=2cm,bottom=2cm,left=3cm,right=3cm,marginparwidth=1.75cm]{geometry}

\usepackage{amsmath}
\usepackage{graphicx}
\usepackage[table]{xcolor}
\usepackage[colorlinks=true, allcolors=blue]{hyperref}

\journal{Journal}

\begin{document}

\begin{frontmatter}



\title{The role of coal plant retrofitting strategies in developing India's net-zero power system: a data-driven sub-national analysis}


\author[inst1]{Yifu Ding}

\affiliation[inst1]{organization={MIT Energy Initiative},%Department and Organization
            addressline={50 Ames St}, 
            city={Cambridge},
            postcode={02142}, 
            state={Massachusetts},
            country={United States}}

\author[inst2]{Dharik Mallapragada}
\author[inst1]{Robert James Stoner}

\affiliation[inst2]{organization={Department of Chemical and Biomolecular Engineering, Tandon School of Engineering, New York University},%Department and Organization
            addressline={726 Broadway}, 
            city={New York},
            postcode={10003}, 
            state={New York State},
            country={United States}}

\begin{abstract}
India set two Nationally Determined Contribution targets to achieve the net-zero carbon emission goal by 2070, which requires deep decarbonization of India’s power generation sector. Yet, coal power generation contributes to more than 60\% of its total power generation, and policies still permit further coal fleet expansion and lifetime extensions. In this paper, we investigate the role of retrofitting India’s coal plants for carbon capture and storage and biomass co-firing in developing the net-zero power system. We model the power generation and transmission network expansions across 30 Indian states in four representative grid evolution scenarios under progressively tighter carbon emission caps, taking into account sub-national coal price variation and thermal efficiency of individual coal plants. We find that coal plant retrofitting could happen by 2035 when an annual carbon cap for the power generation sector is less than 1,000 million tons CO$_2$. This reduces the unabated coal plant capacity, electricity generation, and carbon abatement costs. Exploiting renewable energy potentials solely, such as wind resources, could reduce carbon abatement costs significantly but will result in low coal plant utilization and uneven renewable generation deployment between Southern and Central regions concerning energy justice.

\end{abstract}


\begin{highlights}

\item A data-driven capacity expansion model for India’s net-zero power system is built to explore retrofitting strategies for 806 coal plants  by 2035.

\item Four grid evolution scenarios are considered to model carbon capture and storage, biomass co-firing, and renewable energy integration.

\item Coal plants with carbon capture could happen by 2035 when the carbon cap for power generation is less than 1,000 million tons.

\item Renewable energy integration is cost-effective but could result in low utilization of coal plants and energy injustice issues.

\end{highlights}


\end{frontmatter}


\section{Introduction}

India is the third largest carbon emitter globally, with power generation accounting for over half of the total \cite{iea_country_2022, metcalf_carbon_2021}. India's energy demand is projected to double or triple by 2040 relative to the 2017 level \cite{iea_india_2021}. Endeavoring to meet the global goal of holding warming well below 2 degrees relative to the pre-industrial level, India has committed to net-zero carbon emissions by 2070. Its Nationally Determined Contribution (NDC) targets include installing 500 GW of non-fossil-fuel energy generation capacity and reducing its economy-wide emission intensity by up to 45\% by 2030 relative to 2005 level \cite{ministry_of_power_government_of_india_india_2022}.

India’s existing power generation heavily depends on high-emission coal plants. The total coal plant capacity reached 205 GW and generated more than 60\% electricity generation in 2022 \cite{central_electricity_authority_national_nodate}. The national government has not proposed comprehensive phase-out plans for coal plants to date, but rather will continue to permit new construction and lifetime extensions of existing coal plants \cite{ministry_of_power_phasing_2023}. The decommission of unabated coal plants appears financially and technically infeasible. - This will lead to a great number of coal plants and mines closure with ensuing unemployment in coal-bearing regions \cite{auger_future_2021}, along with declining firm power generating capacity in the face of the increasing summer peak electricity demand \cite{ministry_of_power_phasing_2023, sudarshanvaradhan_india_2023}. These concerns have tended to slow the rate of decommissioning of old plants, especially in low-cost coal regions in the east, thereby perpetuating the carbon emission burden \cite{oskarsson_indias_2021}.

Previous studies \cite{nikit_abhyankar_least-cost_nodate, amy_rose_least-cost_2020, central_electricity_authority_draft_2023, lu_indias_2020, deshmukh_least-cost_2021, rudnick_decarbonization_2022, barbar_impact_2023} have assessed India’s least-cost power system expansion in the next few decades until 2050, mainly focusing on the role of energy storage, renewable generation, and transmission capacity for carbon emission reductions. Nevertheless, these studies show coal plants will remain the main energy source and reserves by 2030. Barbar et al. \cite{barbar_impact_2023} investigate the impact of increasing electricity demand on power system evolution from 2020 to 2050 based on a five-zone India power system. They conclude that deep decarbonization of India’s power sector will require policy measures targeting the existing coal fleet and policies related to accelerating renewable deployment and demand-side reduction. Rose et al. \cite{amy_rose_least-cost_2020} show that battery storage plays a crucial role in renewable generation integration, but supercritical coal plants will be built and concentrated in the east, which accounts for over 90\% of the total carbon emissions by 2047. Lu et al. \cite{lu_indias_2020} models India's power generation mix to achieve 80\% of renewable penetration by 2040. Such a system requires a significant investment in energy storage and network expansion, and over 200 GW of coal plant fleet remains. 

Another research focus is effective carbon policies and renewable energy targets to decarbonize India's power system. Deshmukh et al. \cite{deshmukh_least-cost_2021} model India's power system expansion in 2030 under varying renewable generation targets between 200 to 600 GW, with renewable projects selected based on their levelized electricity generation costs. They conclude that achieving high renewable energy targets will not avert the need to build coal plants. The avoided fossil fuel capacity and carbon abatement cost also change greatly with different shares of wind, solar, and renewable resource quality. Rudnick et al. \cite{rudnick_decarbonization_2022} uses a five-zone India's power system model to optimize the power system planning in 2040 under tradable CO$_2$ emission limits and renewable portfolio standards. Their results show that the tradable CO$_2$ emission limits result in a lower average CO$_2$ abatement cost than are achieved under a renewable portfolio standard.

All these studies above have not considered the role of coal plant retrofits in decarbonizing India's power system, even though the retrofitted coal plants as the low-carbon, firm generation resources could significantly contain carbon abatement costs, such as in the U.S. \cite{sepulveda_role_2018} and China \cite{fan_co-firing_2023}. Coal plants can be retrofitted in several ways to be compatible with a renewable dominant power system, such as by adding carbon capture and storage (CCS) or via fuel-switching strategies involving the use of biofuels or carbon-free fuels such as ammonia and hydrogen. Each of these approaches has different investment requirements within the plant boundary and different supply chain infrastructure needs. For example, CCS retrofits will reduce plant emissions of CO$_2$ and air pollutants (SO$_x$, NO$_x$) but come with substantial capital investment and energy penalty, along with requiring access to CO$_2$ storage reservoirs. By some estimates, India has abundant carbon storage capacity across the nation \cite{vishal_understanding_2021}, with an implied carbon reduction potential of approximately 715 million tonnes per year through CCS \cite{lau_contribution_2023}. The latest Indian government net-zero roadmap also shows that the coal plant with CCS will be implemented between 2030 and 2040 \cite{amit_garg_synchronizing_2024}. Using biomass or green ammonia to replace part of coal as co-firing coal plants appears technically feasible \cite{fan_co-firing_2023, cesaro_ammonia_2021}, but each approach has substantial cost and supply chain-related constraints. For example, the relatively high cost of green ammonia, which is estimated as over \$120/MWh \cite{deng_decarbonizing_2024}, could make its usage as the power generation fuel challenging. In the case of biomass, the availability of biomass stock and the extent of co-firing are key constraints limiting wide-scale adoption. Here, we focus on biomass co-firing in light of its favorable present-day economics compared to co-firing with low-carbon ammonia. Biomass co-firing involves lower capital investment costs with only modest plant modification needed for mixtures of less than 30\% biomass \cite{zhang_understanding_2022}. India already mandated part of coal plants to conduct the co-firing of biomass pellets with a minimum of 5\% fuel mix percentage \cite{ministry_of_power_revised_2023}. 

A few supportive policy measures to reduce carbon emissions or air pollutants of coal power generation have been proposed, including installing flue gas desulfurisation units to reduce air pollutants \cite{us__international_trade_administration_india_2020} and a low tax to coal electricity generation \cite{international_institute_for_sustainable_development_evolution_2020}, but their implementations are far below expectation due to lack of clear sub-national targets. Most literature studying Indian power system planning use the homogeneous coal price and plant characteristics and overlook the sub-national implications \cite{nikit_abhyankar_least-cost_nodate, amy_rose_least-cost_2020, central_electricity_authority_draft_2023, lu_indias_2020, deshmukh_least-cost_2021, rudnick_decarbonization_2022}. Based on life cycle analysis \cite{mallapragada_life_2019}, Indian coal plant carbon emissions has a wide distribution due to inefficient boiler design, indicating the replacement of subcritical coal plants with high-efficiency supercritical coal power plants. With a granular, state-wise power system dispatch model using the empirically derived thermal efficiency curve for coal plants, Sengupta et al. \cite{sengupta_subnational_2022} showed that a more substantial national carbon tax would disproportionately increase the cost to the poorer, coal-heavy eastern states. To understand sub-national goals and circumstances across India, we undertake a geographically granular analysis of the power sector that considers individual coal plant operating characteristics and retrofit opportunities.

Our research uses a power system capacity expansion model (CEM) to investigate the role of coal plant retrofitting strategies in India's power system under various technology and policy scenarios. We develop a 30-state data set for the Indian power system with a reduced representation of inter-state and inter-region transmission networks to optimize future Indian power systems considering two coal plant retrofitting technologies, CCS and biomass co-firing. To precisely model sub-national coal plant retrofitting strategies, we leverage the machine learning clustering techniques to characterize heterogeneous unit-level features of the entire coal fleet. This paper aims to answer several questions: \textit{What is the value of coal plant retrofitting in achieving India's net-zero power system, compared to renewable capacity expansion? What are the sub-national impacts of coal plant retrofitting strategies under various technology and policy scenarios and given heterogeneous unit-level coal plant characteristics? How does the value of these strategies change under different technology and policy scenarios, such as under different renewable capacity limits and transmission network expansion?}

\section{Methods}

\subsection{Capturing heterogeneous characteristics of coal plant units using machine learning clustering}

\begin{figure}[!h]
          \centering
         \includegraphics[width=5in]{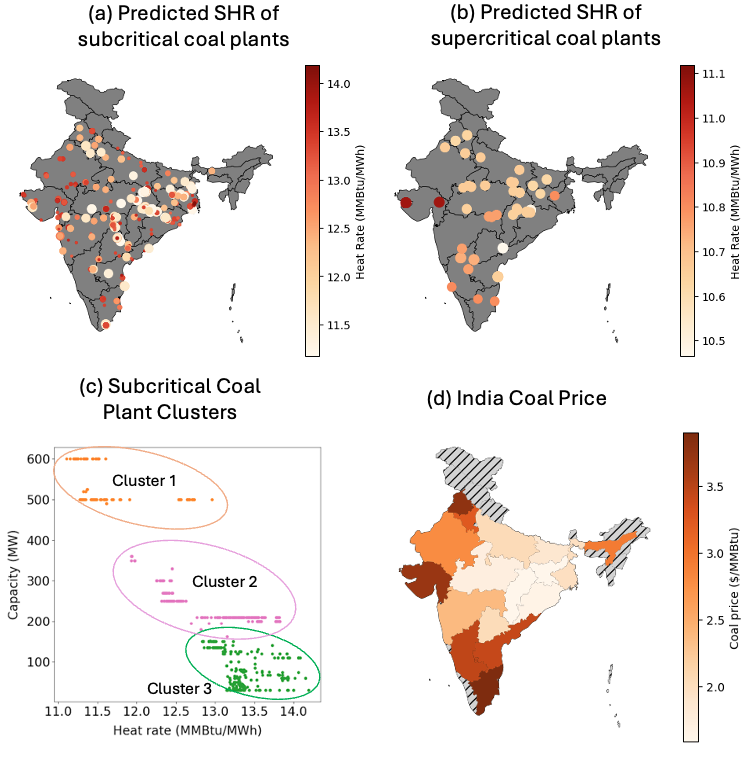}
        \caption{Spatially resolved characteristics of Indian coal plants considered in the study: (a) the predicted SHR of subcritical coal plants, (b) the predicted SHR of supercritical coal plants, (c) Subcritical coal plant clusters, and (d) Spatial distribution in delivered coal price for each state. The shaded areas show the states that currently have no coal plants. All costs are based on the currency rate of 1 INR to 0.012 dollars unless specified otherwise}
         \label{prediction_results}
\end{figure}

We model India's coal fleet as 806 coal-fired power units from our prior work \cite{yifu_ding_dataset_nodate}. The total power capacity is 226 GW, comprising 157 GW from 704 subcritical units and 69 GW from 102 supercritical units \cite{global_energy_monitor_global_nodate}. We consider heterogeneous unit-level characteristics, including station heat rates (SHR), power plant age, and power capacity. 

The SHR represents the thermal efficiency of a power plant, which is defined as the ratio of the heat input to a power plant to the electricity generated by this plant. In our prior work, \cite{yifu_ding_dataset_nodate}, we used the operating SHR measurements for 194 GW of coal plants in 2020 \cite{karthik_ganesan_coal_2021} to predict the operating SHR of 806 India's coal plants specific to two boiler designs, subcritical and supercritical units. Predictions indicate that subcritical coal plants will have SHR values ranging from 10.32 to 14.96 MMBtu/MWh (Figure \ref{prediction_results} (a)), while supercritical coal plants will range from 9.85 to 11.77 MMBtu/MWh (Figure \ref{prediction_results} (b)). The SHR distribution among subcritical coal plants is notably wider compared to supercritical plants, with many subcritical plants exhibiting very high SHR values, indicating a lower thermal efficiency \cite{mallapragada_life_2019}.

To facilitate the computationally tractable evaluation of coal fleet evolution via the CEM, we grouped the 704 subcritical coal plants into four clusters based on their SHRs and power capacity using the K-means clustering algorithm, as depicted in Figure \ref{prediction_results} (c). Subsequently, we inferred the power capacity of coal plants within each cluster for 30 states in India. The rest of the 102 supercritical coal plants were grouped directly by the state, and the representative SHR value and age are the capacity-weighted average values of each state, as presented in \textit{SI, Appendix A, Figure A3}. 

The heterogeneous coal price from ref. \cite{karthik_ganesan_coal_2021} is also incorporated, as shown in Figure \ref{prediction_results} (d). The price includes the coal production and transportation costs, and the shaded area indicates no existing coal plants. The coal price ranges from \$1.59/MMBtu to \$3.90/MMBtu, significantly lower than the natural gas price in India around \$15/MMBtu in 2022. The eastern region has a much lower coal price than the rest of the country since most coal mines are located in these regions, and, therefore, coal from these regions incur lower transportation costs.

\subsection{Capacity Expansion considering Coal Plant Retrofitting}

\begin{figure}[!h]
          \centering
         \includegraphics[width=5in]{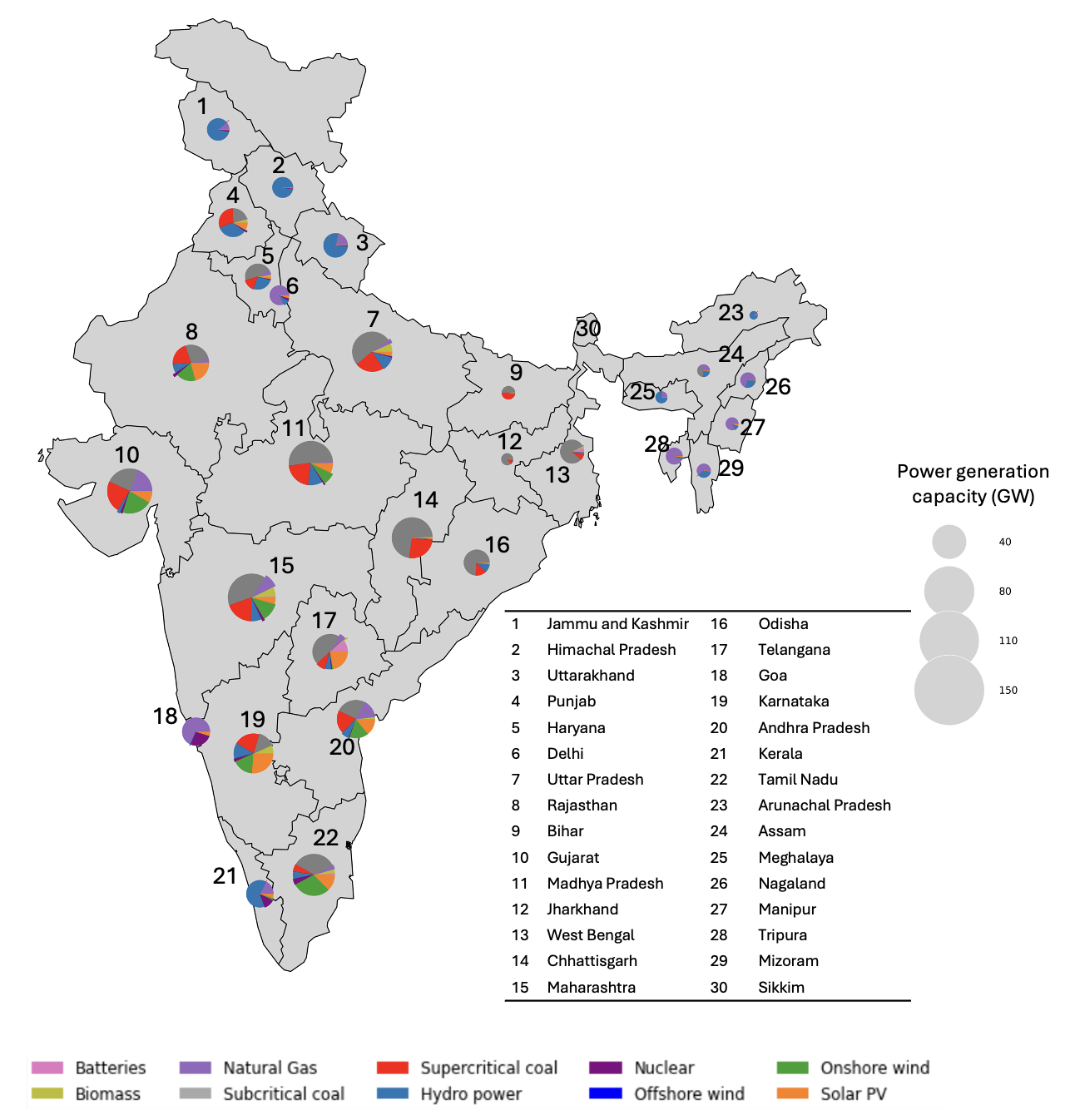}
        \caption{Modelled India 30-state power generation capacity in 2020; The pie chart in each state shows the breakdown of different generation mix.}
         \label{2020_Indian}
\end{figure}

We build a 30-state India's power system planning model based on the open-source CEM, GenX, to co-optimize the power generation and transmission systems \cite{noauthor_genx_nodate}. A brownfield optimization is conducted based on the existing power generation capacity in 2020 and plans for future power generation capacity by 2035. We cluster the hourly renewable and projected electricity demand profiles in 2035 into seven representative weeks, capturing summer and autumn peaks (\textit{SI, Appendix B, Figure B.2}). The 2035 demand scenario accounts for economic-development-driven demand and electricity demand resulting from air conditioning adoption in the building sector \cite{barbar_scenarios_2021}. As presented in Fig. \ref{2020_Indian}, we validated our model by simulating the 2020 power system and comparing it to the actual generation mix. The modeled results show the high accuracy in replicating the power generation mix in the 2020 India power system (\textit{SI, Appendix B, Figure B.8}).

\begin{figure}[!h]
          \centering
         \includegraphics[width=6in]{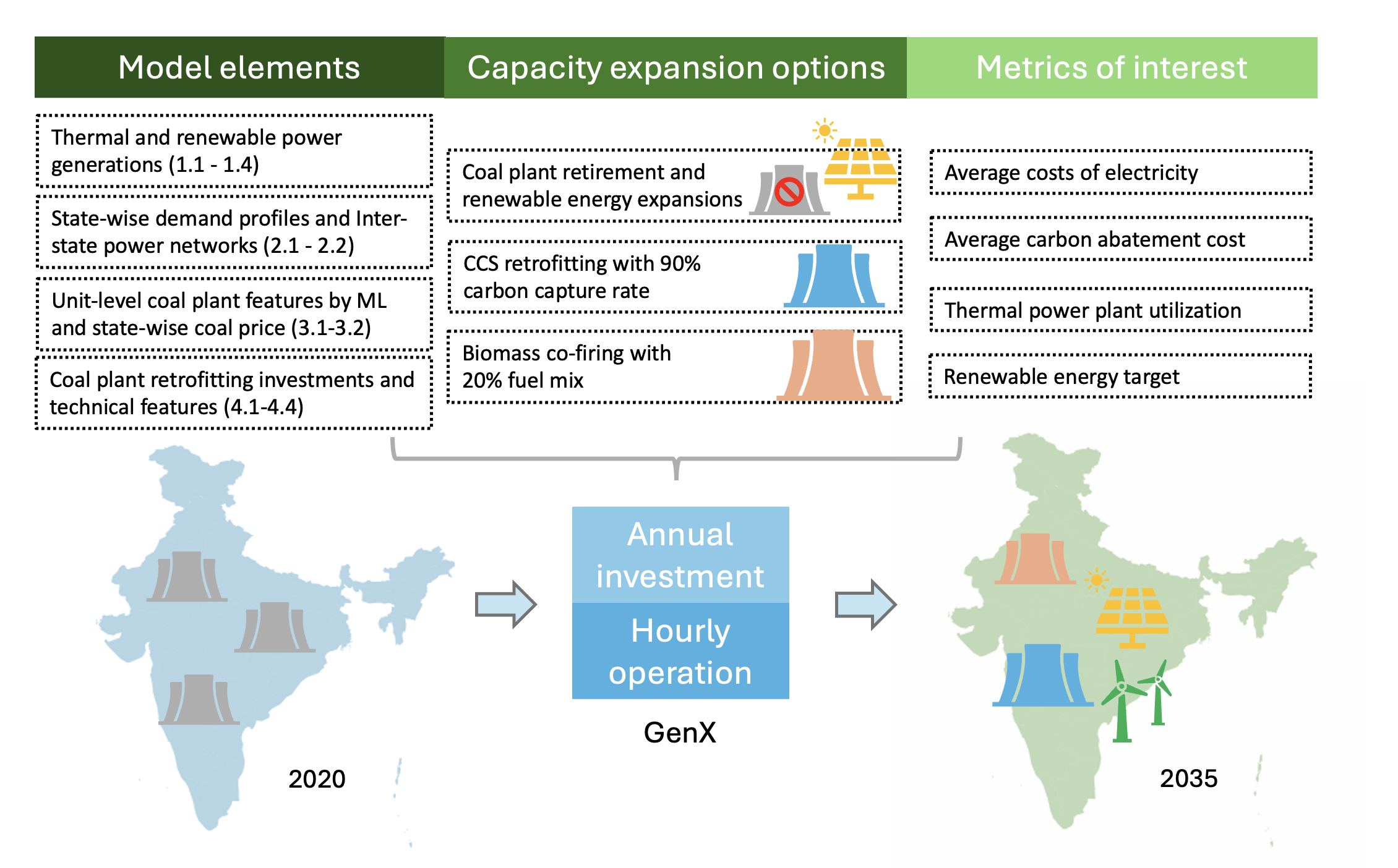}
        \caption{Model framework including model elements, capacity expansion options, and metrics of interests to examine the role of coal plant retrofitting strategies in achieving India's NDC targets; The number in brackets of the model elements refer to the data sources in Table 1}
         \label{model_framework}
\end{figure}

Figure \ref{model_framework} depicts the main framework of the CEM, and Table \ref{data source} summarizes the main data sources. The model incorporates renewable energy resources, thermal power plants, and energy storage, all grouped in 30 states. Renewable energy resources include utility-scale solar PV, hydropower stations, and onshore and offshore wind farms. Thermal power plants include natural gas, sub- and super-critical coal, biomass, and nuclear power plants. We model the linear unit commitment of thermal power plants considering a start-up, ramp-up, and --down. The investment costs of power generation capacity and their technical assumptions are based on India's Central Electricity Authority (CEA) estimations in year 2030 \cite{central_electricity_authority_indian_nodate}. We assume that biomass power plants' capacity will not expand, given the relatively low thermal efficiency. Instead, we consider using biomass co-firing in existing coal power plants. The investments in hydropower and nuclear stations are decided by the government plan \cite{central_electricity_authority_national_nodate}. Our model includes the existing inter-state AC transmission capacity and inter-regional high-voltage DC (HVDC) transmission lines, as presented in \textit{SI, Appendix B, Figure B.7}. We consider inter-state power flows and AC transmission capacity expansion but do allow for investment in any new power transmission path or expanding inter-regional HVDC lines. The relevant data is provided in \textit{SI, Appendix B}.

\begin{table}[!h]
     \centering
     \small
\begin{tabular}{|p{0.5cm} |p{9cm}|p{4.7cm}|}
\hline
\multicolumn{3}{|c|}{\cellcolor{green!25}Thermal and renewable power generations} \\\hline
\textbf{No.} &\textbf{Descriptions} & \textbf{Data Sources} \\ \hline 
1.1&Existing renewable energy, thermal power generation and hydropower station capacity in 2020 & CEA \cite{central_electricity_authority_installed_2019} and India Ministry of new and renewable energy \cite{ministry_of_new_and_renewable_energy_renewable_2022} \\ \hline
1.2&The planned capacity of nuclear power stations and hydropower plants by 2035 & CEA \cite{central_electricity_authority_national_nodate} \\ \hline
1.3&Hourly capacity factor profiles of hydropower station, renewable energy resources, and geospatial renewable potentials & NREL ReEDS - India \cite{nrel_regional_nodate}; MERIT India \cite{noauthor_csep_nodate} \\ \hline
1.4&Overnight investment costs and technical parameters of thermal power generations (except coal power plants)  & CEA \cite{central_electricity_authority_indian_nodate}   \\ \hline
\multicolumn{3}{|c|}{\cellcolor{cyan!25}State-wise demand profiles and inter-state power networks} \\\hline
2.1&Projected electricity demand profiles of 30 Indian states in 2035 considering electric vehicles and air conditioners & Barbar et al. \cite{barbar_scenarios_2021} \\ \hline
2.2&Inter-region and inter-state power network topology and transmission capacity  & Indian Ministry of Power \cite{ministry_of_power_government_of_india_power_2023, noauthor_changing_nodate}, and NREL ReEDS - India \cite{nrel_regional_nodate}    \\ \hline
\multicolumn{3}{|c|}{\cellcolor{orange!25} Unit-level coal plant features and state-wise coal price} \\\hline
3.1& Operational SHR value of 806 India coal-fired power units & Methods in Section 2 \cite{yifu_ding_dataset_nodate} \\ \hline
3.2& State-wise coal price  & Ganesan et al. \cite{karthik_ganesan_coal_2021}  \\ \hline
\multicolumn{3}{|c|}{\cellcolor{pink!25} Coal plant retrofitting investments and technical features} \\\hline
4.1&Thermal efficiency and power capacity penalty factors for the CCS and Biomass co-firing retrofitting & NREL Electricity Annual Technology Baseline \cite{nrel_electricity_2023} \\ \hline
4.2&Overnight investment and operational costs for CCS and biomass co-firing coal plant retrofitting & IRENA \cite{irena_renewable_2021}; Fan et al. \cite{fan_co-firing_2023}\\ \hline
4.3&Estimated costs of carbon transportation and storage in the 30 Indian states & Lau \cite{lau_contribution_2023}; Vishal et al. \cite{vishal_understanding_2021}\\ \hline
4.4&Biomass power generation potentials in the 30 Indian states & Indian ministry of new and renewable \cite{ministry_of_new_and_renewable_energy_national_2023} \\
\hline
\end{tabular}
\caption{Summary of the main data sources used in this study}
 \label{data source}
\end{table}

The model considers investment in supercritical coal plants and retiring or retrofitting existing coal plants. Based on India's policy as of 2023 \cite{ministry_of_power_phasing_2023}, we assume that no coal plants will be retired even if their age exceeds 50 years by 2035. Instead, they will still be used after renovation and life extension at the same SHR. The CCS retrofitting option is only considered for supercritical coal plants with a large power capacity ($>$ 500 MW$_e$) \cite{lau_contribution_2023}, and therefore 69 GW of existing super-critical coal plants is considered to be feasible for CCS by 2035. The power capacity reduction and additional fuel consumption due to the carbon capture process are modeled using the power capacity and thermal efficiency penalty factors, respectively (\textit{SI, Appendix C, Table C.1}). We estimated the incurred CO$_2$ transportation and storage costs in each state, which is translated into an effective fuel cost penalty, ranging from \$0.8 - \$2.4/MMBtu (\textit{SI, Appendix C, Figure C.5}). Biomass co-firing retrofitting can be conducted on any coal power unit, and the thermal efficiency is assumed to be unchanged after retrofitting due to the boiler renovation. In other words, zero thermal efficiency and power capacity penalty factor. We set the maximum power capacity limits for the planned biomass co-firing coal plants of each state as the biomass power generation potentials published by Indian government \cite{ministry_of_new_and_renewable_energy_national_2023} (\textit{SI, Appendix C, Figure C.6}).

\subsection{Renewable generation expansion and coal plant retrofitting scenarios for India's grid decarbonization evolution}

Four scenarios for the least-cost power system expansion are evaluated. They are scenarios of `baseline,' `high renewable capacity,' `CCS \& Biomass co-firing', and `biomass co-firing only.' We analyze India's grid decarbonization evolution in these four scenarios under four different CO$_2$ emissions scenarios. The carbon intensity of India's power generation in 2005 was approximately 901.7 kgCO$_2$/MWh \cite{shearer_future_2017}. We assume a 45\% reduction of carbon intensity in the power generation sector in line with NDC targets, and this gives a carbon intensity of 495.9 kg CO$_2$/MWh and a total carbon emission of power generation of around 1,130 Mt CO$_2$ in 2030, which are consistent with the CEA estimation \cite{central_electricity_authority_draft_2023}. The Indian government estimated that this carbon emission will peak from 2035 to 2040 \cite{amit_garg_synchronizing_2024}. We set a 1,000 Mt carbon cap scenario to meet the NDC target, and the 800 and 500 Mt CO$_2$ cap scenarios represent two stringent carbon reduction policies beyond this target.

\begin{figure}[!h]
          \centering
         \includegraphics[width=6in]{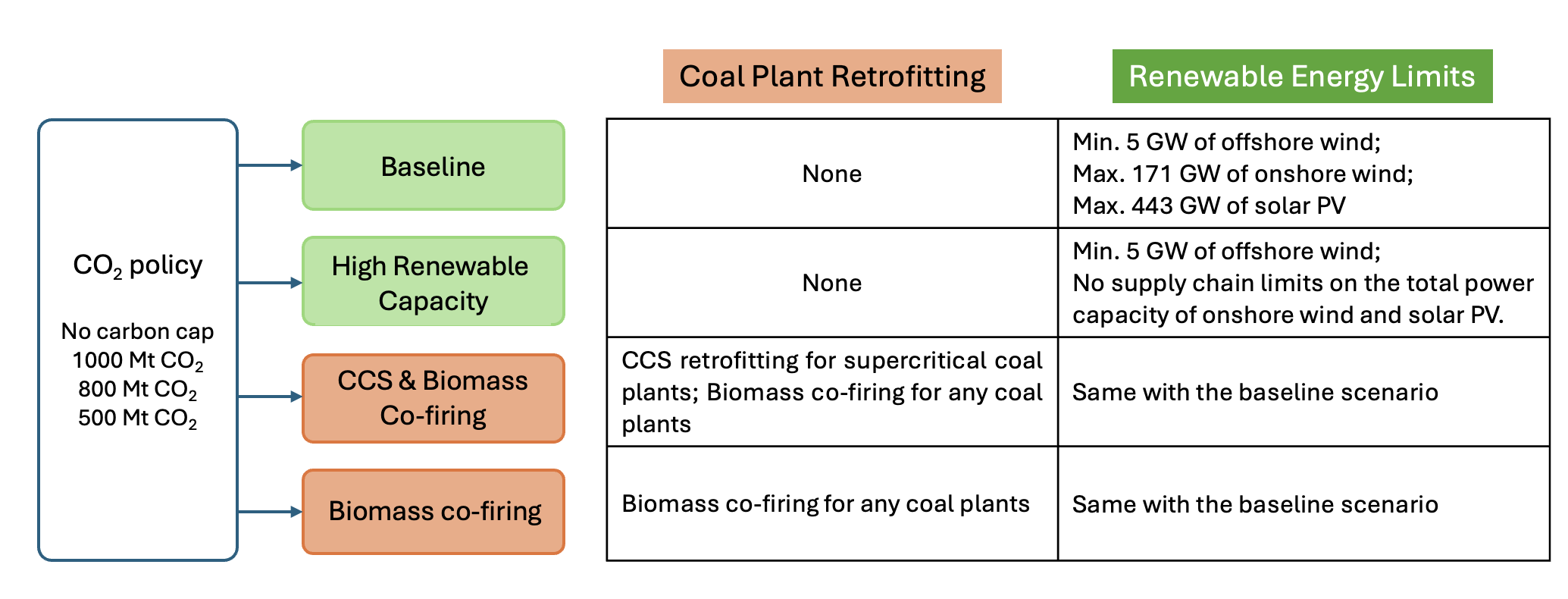}
        \caption{The settings of four renewable generation expansion and coal plant retrofitting scenarios considering India's grid decarbonization evolution}
         \label{scenario_plots}
\end{figure}

Figure \ref{scenario_plots} illustrates that the baseline and high renewable capacity scenarios focus on renewable energy capacity expansion for decarbonization. While the renewable energy potentials are decided by resource quality and land use \cite{nrel_regional_nodate, von_krauland_india_2024}, we set 171 GW and 443 GW as the maximum power capacity limits for the onshore wind and solar PV respectively for the baseline scenario, which are based on India's national renewable generation targets \cite{global_energy_wind_council_accelerating_2022}. The baseline scenario also assumes that India will build at least 5 GW of offshore wind farms in the next ten years \cite{ministry_of_power_government_of_india_zero_2017} with further expansion allowed if deemed economical. To explore the impacts of renewable energy potentials, we create the high renewable capacity scenario with an unlimited solar and wind supply chain, which means there is no limitation on the total onshore wind and solar PV capacity installed. The high renewable capacity scenario also has a more ambitious target to construct at least 30 GW offshore wind capacity by 2035. 

Two other coal plant retrofitting scenarios explore another dimension of decarbonizing India's power system. India has yet to equip any coal plant with the CCS, and the construction of the CCS facility is uncertain due to the high capital cost and social acceptance. We consider two coal plant retrofitting scenarios with and without the CCS infrastructure. The `CCS \& Biomass co-firing' allows for the option to retrofit supercritical coal plants with CCS and the option that any coal plant can be retrofitted into biomass co-firing power plants with 20\% biomass co-firing. In contrast, the `biomass co-firing only' scenario does not allow CCS coal plant retrofitting.

\section{Results}

\subsection{Capacity expansion and power generations}

\begin{figure}[!h]
          \centering
         \includegraphics[width=6in]{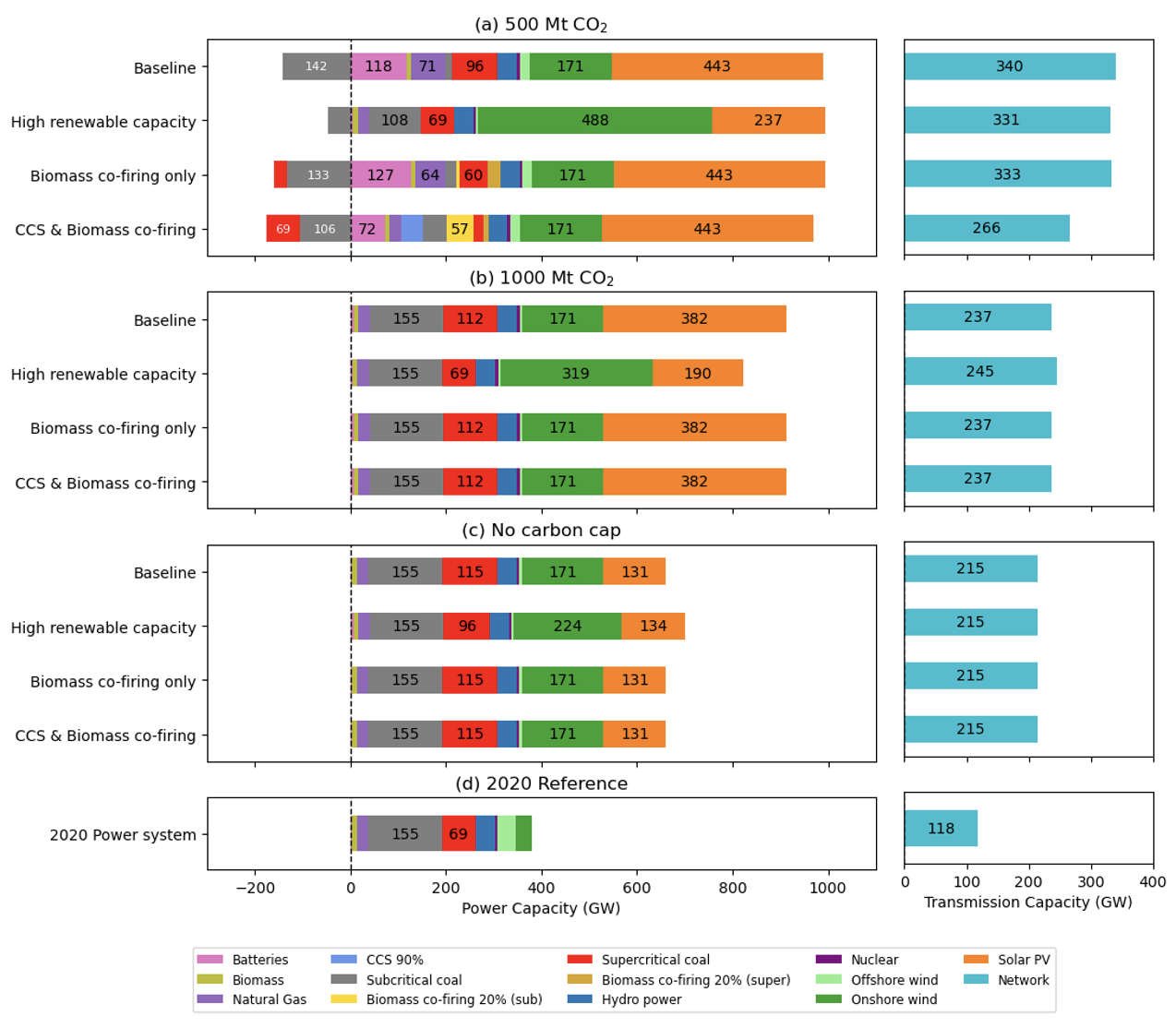}
        \caption{India's power generation capacity considering the retired or retrofitted coal plants and transmission network power capacity by 2035 under (a) 500 Mt CO$_2$ cap, (b) 1,000 Mt  CO$_2$ and (c) no carbon cap in four scenarios: Baseline, High Renewable Capacity, Biomass co-firing only, CCS \& Biomass co-firing; The right panel shows India's power generation capacity considering the retired or retrofitted coal plants and the right panel shows the power transmission capacity. The values are labeled for power capacity higher than 50 GW.}
         \label{capacity_gen_mix}
\end{figure}

Figure \ref{capacity_gen_mix} (a) - (d) shows the planned generation and transmission capacity by 2035 in four scenarios, under 500, 1,000 Mt CO$_2$ cap, and no CO$_2$ cap, respectively.  It reveals India could achieve the non-fossil-fuel energy capacity target of 500 GW without stringent carbon policy support. In the high renewable capacity scenario without a carbon cap, the total renewable generation capacity (including hydropower stations) reaches 413 GW. The installed renewable generation capacity exceeds 500 GW when a moderate annual carbon cap of less than 1,000 Mt CO$_2$ is enforced, leading to the deployment of 171 GW of onshore wind, 5 GW of offshore wind, 40.5 GW of hydropower, and 382 GW of solar PV. However, the deployment of energy storage and natural gas power plants is limited even in cases with an increasingly stringent carbon cap.

The onshore wind resource plays an important role in the deep decarbonization of India's power system. Our results show that the onshore wind supply chain limit is reached in three scenarios except the high renewable capacity scenario. Even without the supply chain limit, the planned onshore wind capacity is far less than the actual wind power potentials published by CEA (i.e., 302 GW at the height of 100 meters and 695.5 GW at 120 meters) \cite{central_electricity_authority_national_nodate}. A substantial transmission network expansion is required to support power system decarbonization. India had approximately 118 GW of inter-state AC transmission capacity by the end of 2020 \cite{noauthor_india_2023, iea_india_2021}. An additional transmission capacity of 106 - 222 GW is expected to be built across all scenarios, and the planned transmission power capacity increases with a more stringent carbon cap.

Early coal plant retirement and retrofitting will only happen under a CO$_2$ cap of less than 1,000 Mt. Sub-critical coal plants will be early retired when the carbon cap is 1,000 Mt CO$_2$ and completely decommissioned when the carbon cap reduces to 500 Mt CO$_2$ cap. The high renewable capacity scenario will retain more existing sub-critical coal plants for flexibility than the other three scenarios. 
\begin{figure}[!h]
          \centering
         \includegraphics[width=6.3in]{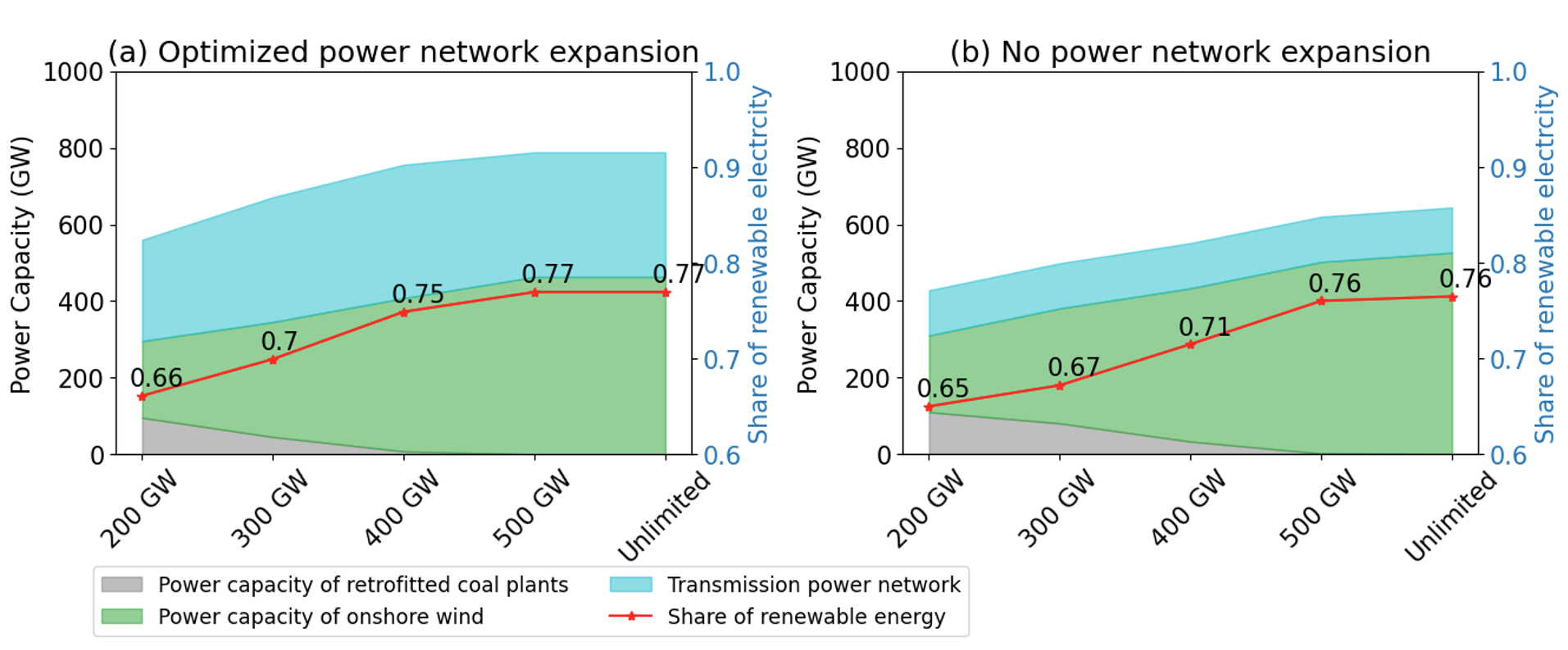}
        \caption{Changes in renewable electricity share, power capacity (GW) of onshore wind and retrofitted coal plants under the increasing onshore wind capacity supply chain limits and 500 Mt CO$_2$ cap in the (a) optimized power network expansion and (b) no power network expansion cases}
         \label{supply_chain_limit_test}
\end{figure}

The CCS \& Biomass co-firing and biomass co-firing only scenarios illustrate the economic feasibility of two retrofitting technologies. In the `CCS \& Biomass co-firing' scenario, 43 GW of supercritical coal plants will be retrofitted into the CCS-coal power plant under 500 Mt CO$_2$ cap, and 66 GW of subcritical coal plants will be retrofitted into biomass co-firing power plants with 20\% fuel mix. This indicates that biomass co-firing with a low biomass fuel mix is a less cost-effective decarbonization approach compared to the CCS retrofitting when implemented on supercritical coal power plants. Likewise, the biomass co-firing only scenario without CCS will only retrofit less than 30 GW of supercritical coal plants into the co-firing power plants under 500 Mt CO$_2$ cap.

 We found that the unlimited renewable generation capacity will lead to no retrofitted coal plants, as illustrated in the additional scenario, CCS \& Biomass co-firing with high renewable capacity. As presented in Figure \ref{supply_chain_limit_test}, we conduct simulations with a progressively increasing onshore wind supply chain limit from 200 to 500 GW under the co-optimized and no power network expansion. The CCS-coal capacity will decrease as the onshore wind capacity expands and diminishes when the onshore wind capacity reaches 400-500 GW. Even under the unlimited renewable supply chain limits and 500 Mt CO$_2$ cap, the share of renewable electricity will plateau at 76.9\%, remaining around 440 TWh of coal power generations (\textit{SI, Appendix D, Figure D.2}).

\subsection{Geographical distributions of renewable power generations, retired and retrofitted coal plants}

\begin{figure}[!h]
          \centering
         \includegraphics[width=6in]{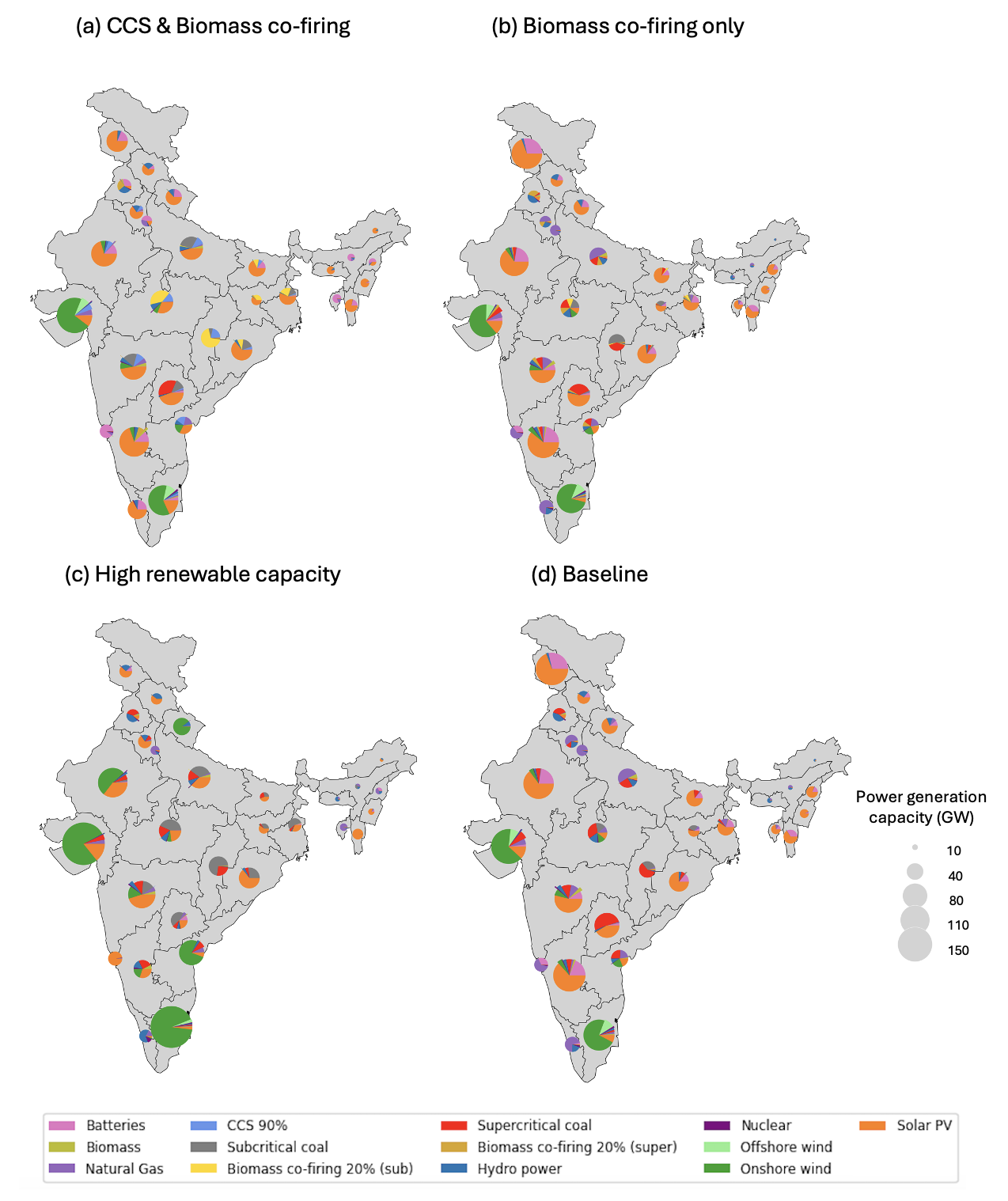}
        \caption{India's state-wise power generation capacity by 2035 under the 500 Mt CO$_2$ cap in the selected four scenarios: (a) CCS \& Biomass co-firing; (b) Biomass co-firing only; (c) High renewable capacity; and (d) Baseline scenarios}
         \label{power_generation_difference}
\end{figure}

Figure \ref{power_generation_difference} depicts the sub-national power generation in these four representative scenarios under 500 Mt CO$_2$ cap. We also compare the sub-national differences in the total generation and transmission capacity under no carbon cap and 500 Mt CO$_2$, as presented in the \textit{SI, Appendix D, Figure D.3}. Our results show that the transmission capacity between the Central state of Madhya Pradesh and its neighboring states reduces substantially when the carbon cap reduces because of reductions in the coal plant capacity in the Central state. The interstate transmissions within regions are reinforced to support renewable resource integration, considering their spatial-temporal variability.

Energy inequality is a historical issue in India's energy system development \cite{bhattacharyya_all_2022}. - India's energy infrastructure is regulated within each state, and most of the renewable generation capacity is found in southern and western India, where wealthier states are located \cite{sengupta_subnational_2022}. Among the four scenarios, the high renewable capacity scenario has the most uneven distribution of renewable generation capacity. The onshore and offshore wind farms are mainly concentrated in the Western and Southern states, including Rajasthan, Tamil Nadu, and Gujarat, alongside a sharp reduction in thermal power generation in eastern and central India. To meet the energy demand, the power network capacity is increased substantially to transmit the excess renewable generation from the western region to the northern and eastern regions, as presented in \textit{SI, Appendix D, Figure D.3 (c)}.

In the CCS \& biomass co-firing scenario, renewable generations are distributed more evenly across 30 Indian states. Batteries and low-carbon and firm generation, such as retrofitted coal plants with CCS and biomass co-firing plants, support the solar power generation integration across India, especially in states such as Madhya Pradesh, Bihar, Jharkhand, and Chhattisgarh. The power generation of biomass co-firing only and the baseline scenario has similar distributions (Figure \ref{power_generation_difference} (b) and (d)). - In both scenarios, low-efficiency sub-critical coal plants are retired, and natural gas power plants are planned for decarbonization, such as in Uttar Pradesh and Delhi.

\begin{figure}[!h]
          \centering
         \includegraphics[width=5.5in]{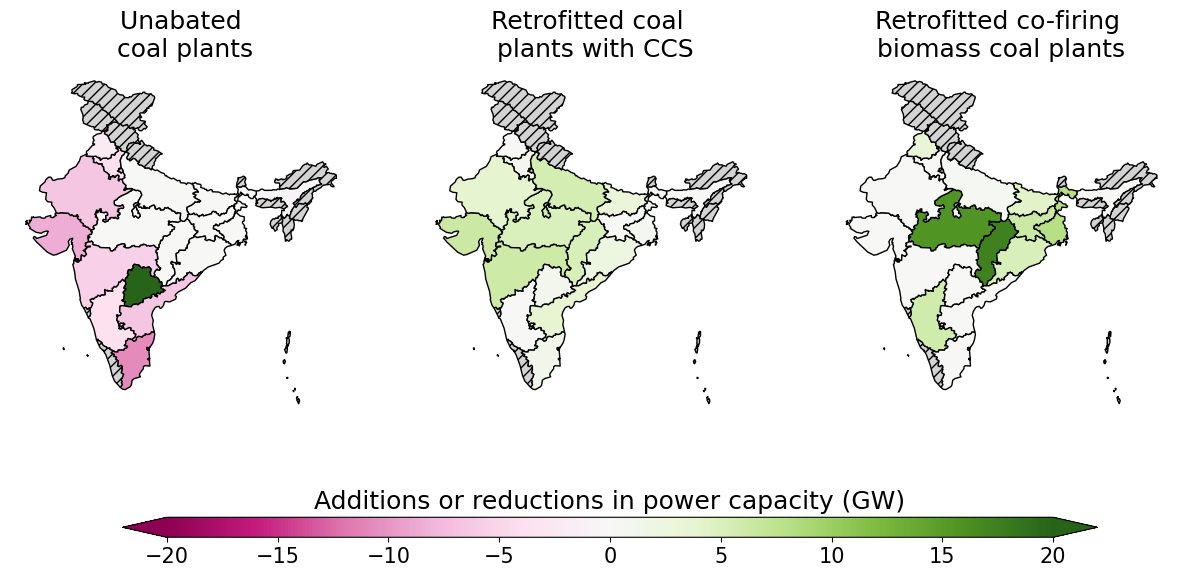}
        \caption{Changes in the unabated coal plants, retrofitted CCS and biomass co-firing coal plants compared with the starting capacity of 2020 India's power system in the `CCS \& Biomass co-firing' scenario under 500 Mt CO$_2$ cap; The shaded areas shows the states that currently have no coal plants}
         \label{capacity_increment}
\end{figure}

Figure \ref{capacity_increment} shows the changes of the unabated coal plants, retrofitted CCS, and biomass coal plants compared relative to their start capacity in the 2020 India's power system for the `CCS \& Biomass co-firing' scenario under 500 Mt CO$_2$ cap. The low-efficiency sub-critical coal plants are retired in Western and Southern states where coal prices are the highest, while new coal plants or retrofitting are planned in the Central and Eastern regions with lower coal prices.

\subsection{Coal plant capacity and utilization}

\begin{figure}[!h]
          \centering
         \includegraphics[width=6in]{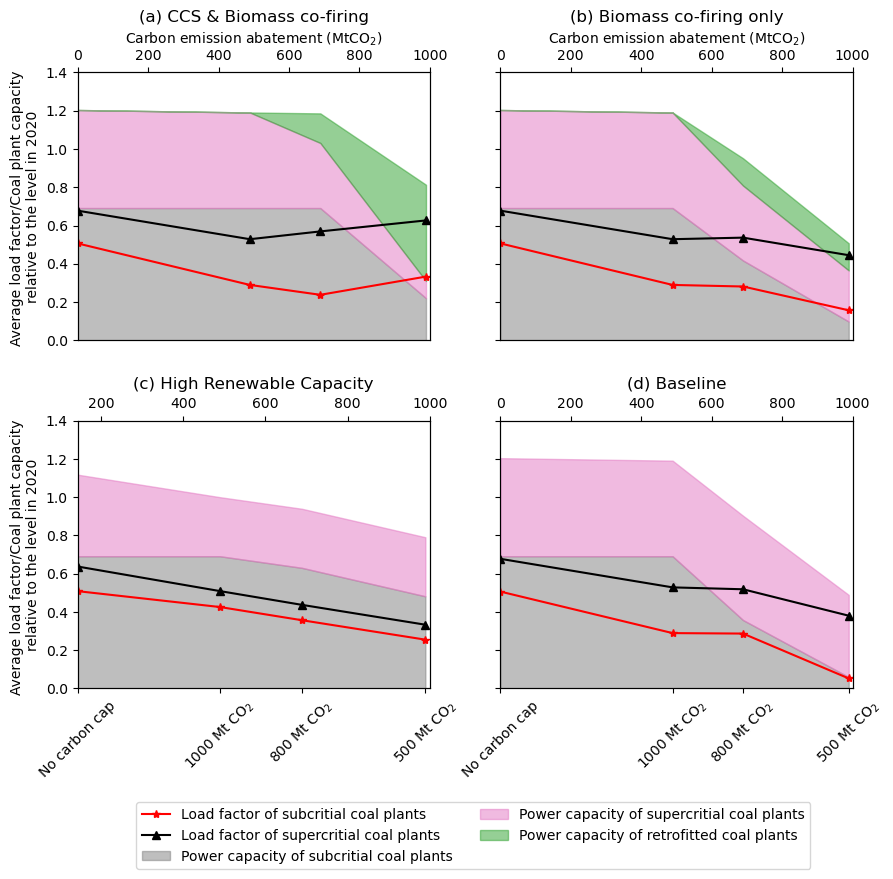}
        \caption{The power capacity of sub- and super-critical coal power plants relative to the stating coal plant capacity in 2020 with the average load factors in the (a) CCS \& Biomass co-firing and (b) Biomass co-firing only (c) High Renewable Capacity, and (d) Baseline scenarios}
         \label{capacity_factor}
\end{figure}

Coal plant retrofitting can change the operating regimes of coal plants, for example, whether they supply the base or peak load. Here, we use the plant load factor, which is defined as the ratio of the actual electricity output to the electricity generated at its full nameplate capacity over a year. A higher plant load factor means better thermal plant utilization. Figure \ref{capacity_factor} (a) - (d) shows the plant load factor for sub-, super-critical, and retrofitted supercritical coal plants and their power capacity under different carbon constraints relative to the starting coal plant capacity in 2020 for the four selected scenarios.

India's coal plant capacity will increase around 1.1-1.2 times by 2035 across four scenarios, driven by the increasing electricity demand. The total coal capacity will reduce when the carbon cap becomes stringent. Under the 500 Mt CO$_2$, the baseline scenario has the lowest coal plant capacity among the four scenarios since almost all the subcritical coal plants are retired. In contrast, the high renewable capacity scenario has the highest coal plant capacity due to the flexibility needed. The CCS \& Biomass co-firing scenario also remains a high coal plant capacity, though half of the coal plant capacity comes from the retrofitted coal plants.

As shown in Figure \ref{capacity_factor} (a) and (b), the average load factor of coal plants also decreases as the carbon cap becomes more stringent, primarily due to the reduced utilization of coal plants in response to higher carbon prices. Supercritical coal plants typically exhibit a higher load factor than subcritical coal plants, often exceeding 50\%, and supply the base load. Under the 500 Mt CO$_2$ cap, in the baseline scenario (d), the load factor of subcritical coal plants drops to below 0.1, which means a number of unabated subcritical coal plants will only operate for a few hours across a year to meet peak demand. coal plant retrofitting could increase the load factors of coal plants and improve thermal plant utilization. - The load factors of subcritical and supercritical coal plants increase when the carbon cap reduces from 800 to 500 Mt in the CCS \& Biomass co-firing scenario.

\begin{figure}[!h]
          \centering
         \includegraphics[width=6in]{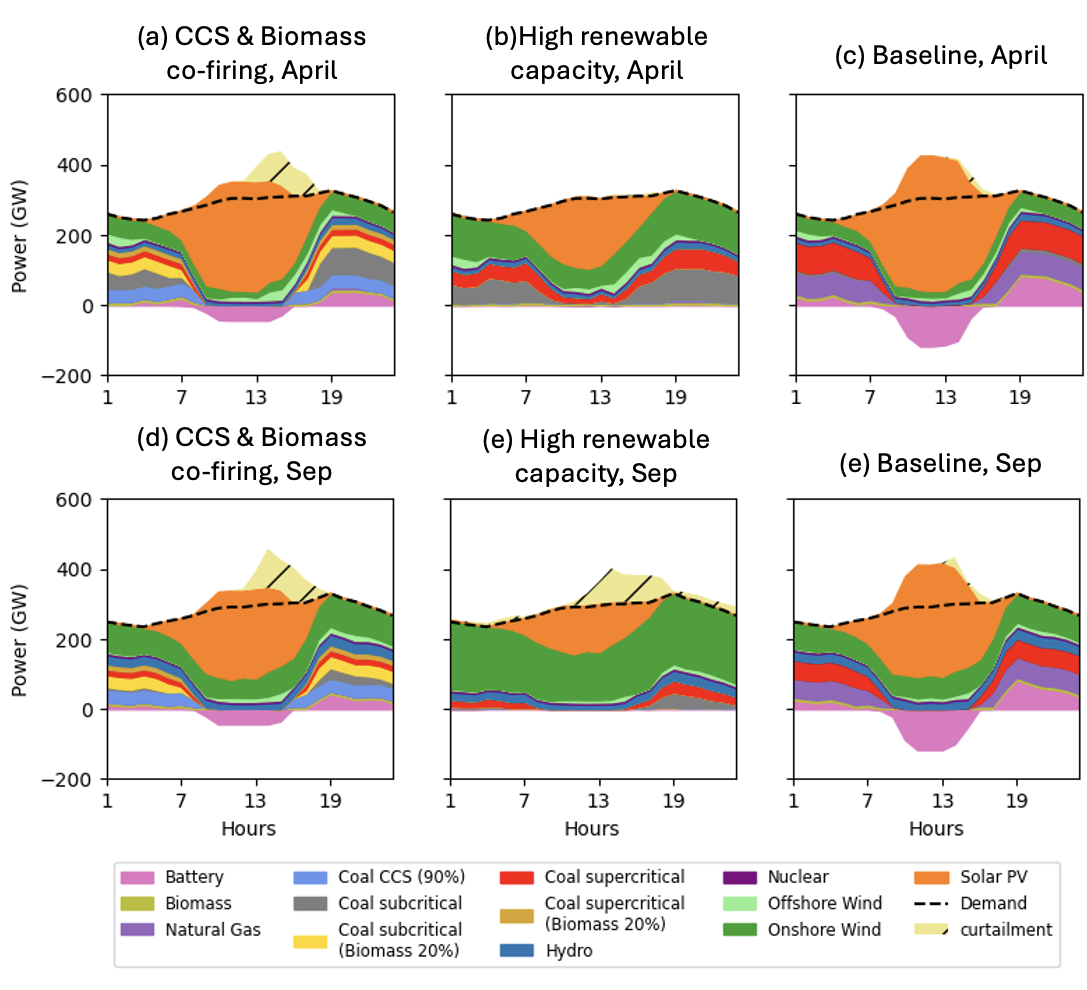}
        \caption{Daily power dispatch for three scenarios in April and September under 500 Mt CO2: (a) CCS \& Biomass co-firing - April, (b) high renewable capacity - April, (c) Baseline - April, (d) CCS \& Biomass co-firing - Sep, (e) High renewable capacity - Sep, and (e) Baseline - Sep}
         \label{dispatch_network}
\end{figure}

Figures \ref{dispatch_network} (a) - (e) show the daily power dispatch for three scenarios, CCS \& Biomass co-firing, High Renewable Capacity, and Baseline in April and September under 500 Mt CO$_2$. In the CCS \& Biomass co-firing and Baseline scenarios, batteries and coal plants serve as the primary flexibility resources to support significant ramping-up and -down up to 40 GW per hour, equivalent to around eight times the magnitude of the California duck curve \cite{us_energy_information_administration_as_nodate}.  In the baseline scenario, subcritical coal plants generate minimal power output only to meet the afternoon peak demand. Batteries and natural gas provide significant flexibility, leading to little renewable curtailment. The daily power dispatch of the biomass co-firing only scenario is similar to the baseline scenario, which is presented in \textit{SI, Appendix D, Figure D.4}. While in the CCS \& Biomass co-firing scenario, the carbon capture procedure involves the additional start-up fuel consumption of 27\% (\textit{SI, Appendix C, Table C.1}), and therefore, the retrofitted supercritical coal plants with CCS change power outputs less flexibly to reduce the start-up cost. This leads to renewable generation curtailments during the afternoon ramp-up. In the high renewable capacity scenario, the onshore wind power generation complements solar power generation in the spring and summer, which leads to a low renewable curtailment (Figure \ref{dispatch_network} (b)), in contrast to a higher renewable curtailment in Sep due to the wind spillage. Notably, the model neglects network congestion, insufficient local power dispatch, and power loss at the distribution network. Thus, renewable curtailment could be significantly higher in reality.

\subsection{Electricity generation and carbon abatement costs}

\begin{figure}[!h]
          \centering
         \includegraphics[width=6in]{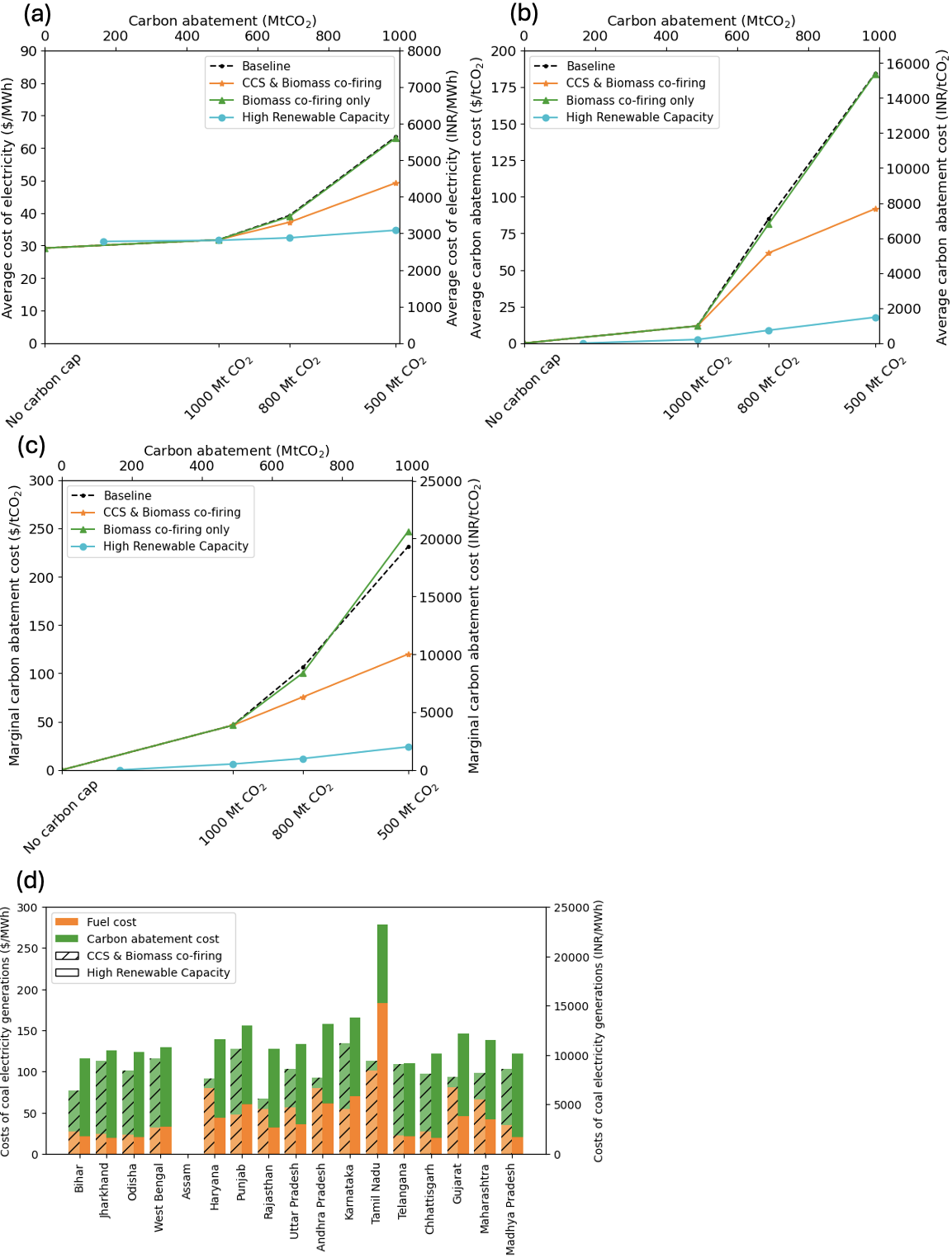}
        \caption{(a) Average cost of electricity, (b) average carbon abatement costs and (c) marginal carbon abatement cost in the baseline, CCS \& Biomass co-firing, Biomass co-firing only, and High renewable capacity scenarios under different carbon caps; (d) the costs of coal electricity generation under 500 Mt CO$_2$ cap}
         \label{system_costs_three_scenario}
\end{figure}

We calculate the average electricity generation by dividing the total system cost (planning and operational costs) with the total energy consumption, and we calculate the average carbon abatement cost by comparing the difference in the total system cost of the selected emission cap case and no emission cap case, divided into the per-unit energy consumption. Figure \ref{system_costs_three_scenario} (a) shows the average cost of electricity under different carbon caps in four scenarios. In the baseline scenario, the average cost of electricity increases from \$29/MWh under no carbon cap to \$63/MWh under a 500 Mt CO$_2$ cap. In the CCS \& Biomass co-firing scenario, this cost reduction in the average cost of electricity is incurred when a carbon cap of less than 1,000 Mt CO$_2$ is enforced. The cost reduction due to the CCS retrofitting mainly results from the reduced fuel consumption costs (Figure \ref{system_costs_three_scenario} (d)). The high renewable capacity scenario has the highest average cost of electricity under no carbon cap, due to high investments of the offshore wind farm. The cost becomes lower than the baseline when the carbon cap becomes less than 1,000 Mt CO$_2$ and increases marginally as the carbon cap becomes more stringent.

The average carbon abatement cost is compared in Figure \ref{system_costs_three_scenario} (b). The average carbon abatement cost is calculated as the average additional system cost per ton incurred for the CO$_2$ emission avoided compared with the no carbon cap case. Notably, the average carbon abatement cost differs from the marginal carbon abatement cost, which sets the carbon price in the market, and we also report it in Figure \ref{system_costs_three_scenario} (c). In the baseline scenario, the average carbon abatement cost will increase rapidly to \$180/tCO$_2$ under the 500 Mt CO$_2$, which exceeds the minimum estimation of the U.S. social cost of carbon (i.e., \$140/tCO$_2$) \cite{us_environmental_protection_agency_epa_2023}. The high renewable capacity sees a greater reduction in carbon abatement costs. Biomass co-firing retrofit only with a low biomass fuel mix will not reduce the electricity generation and carbon abatement costs.

We calculate the fuel and average carbon abatement costs per unit of coal electricity generation in different states where there are coal plants in the CCS \& Biomass co-firing and high renewable capacity scenarios under a 500 Mt CO$_2$ cap. The carbon abatement cost is calculated by multiplying the carbon emission of per-unit coal electricity generation and the average carbon abatement cost. As shown in Figure \ref{system_costs_three_scenario} (c), the cost of coal electricity generation in the high renewable capacity scenario is higher than in the CCS \& Biomass co-firing scenario under a stringent carbon cap, despite the former having a lower average cost of electricity. The increased fuel and carbon abatement costs are due to the low plant utilization and high carbon emission, particularly in the states of Tamil Nadu, Gujarat, and Andhra Pradesh. Assam has no coal electricity generation cost since all coal power plants in this state are retired.

\section{Policy implications and model limitations}

To explore the role of coal plant retrofitting in developing India's net-zero power system, we create a data-driven 30-state Indian power system model and four grid evolution scenarios considering coal plant retrofitting strategies and renewable generation integration. Results show that there is an opportunity for coal plant retrofitting, especially CCS, in developing India's net-zero power system by 2035 since it exhibits several advantages in reducing the unabated coal capacity and improving coal plant utilization. We summarize the following policy implications.

\textit{The CCS implementation in the power generation sector requires additional incentive and carbon policy support:} Supercritical coal plants with CCS will happen by 2035 under a carbon cap of less than 1,000 Mt. Under a carbon cap of 500 Mt, 69 GW of supercritical coal power plants will be retrofitted with CCS, giving a high carbon price (i.e., marginal carbon abatement cost) of \$140/tCO$_2$. Another option is to employ carbon capture storage and utilization (CCUS) in an enhanced oil field such as the Bombay offshore oil field \cite{lau_contribution_2023}. This could incentive CCS by obtaining a return from the increased oil production. 

\textit{Renewable integration should consider energy justice to support the just transition in India:} Fully exploiting renewable generation potentials, especially wind capacity to the supply chain limit, will concentrate generation capacity in the Western and Southern where the wind resource is abundant. The optimal coal plant retrofitting could balance generation capacity distributions to support solar generations, preserving millions of coal-related jobs in the poor, coal-bearing eastern states for a just transition in India. 

\textit{Substantial network expansion is needed for India's net-zero power system development:} Our results show that India needs at least an additional 100 GW transmission network capacity for renewable integration by 2035, excluding HVDC transmission lines. The alternative is to explore the decentralized solution such as solar micro-grid \cite{palit_towards_2022} or demand-side management \cite{shu_coal_2023}. This could effectively reduce transmission power losses and improve energy access in rural areas.

Our modeling assumptions have a few limitations. First, we do not consider domestic biomass transportation and associated carbon emissions. A comprehensive life cycle analysis and detailed biomass transportation model could improve results. Secondly, we do not account for the capital investment for CO$_2$ pipelines because that CO$_2$ transportation involving multiple sectors such as chemical refinery needs a more integrated assessment. This means the costs of electricity generation and carbon abatement in the CCS \& Biomass co-firing scenario could be higher. Finally, the model does not incorporate technical issues in the coal plant retrofitting processes, such as land use and water stress, requiring a detailed examination of each plant.

\section{Acknowledgment}

We acknowledge the funding support from IHI Corporation, Japan. We thank Prof. Gilbert E. Metcalf for comments on the manuscript.

\appendix

\section{Supplement Materials}

The supplementary material can be found in the attached file. The model is built based on GenX version v0.3.6., and the model inputs and original results can be found in \href{https://doi.org/10.5281/zenodo.12684827}{https://doi.org/10.5281/zenodo.12684827}.

\bibliographystyle{elsarticle-num}
\bibliography{sample.bib}

%\printbibliography





% \end{thebibliography}
\end{document}